\documentclass[useAMS,usenatbib]{mnras}
\usepackage{graphicx}
\usepackage{color}
\usepackage{epsfig}
\usepackage{amsmath,amssymb}
\usepackage{natbib}
\usepackage{comment}
\usepackage{multirow}
\usepackage{lscape}
\usepackage{threeparttable}
\usepackage{bm}
\usepackage{ulem}

\def\pddt#1#2{\frac{\partial #1}{\partial #2}}

\def\simge{
    \mathrel{\rlap{\raise 0.511ex
        \hbox{$>$}}{\lower 0.511ex \hbox{$\sim$}}}}
\def\simle{z
    \mathrel{\rlap{\raise 0.511ex
        \hbox{$<$}}{\lower 0.511ex \hbox{$\sim$}}}}

\newcommand{\figref}[1]{Fig.~\ref{#1}}
\newcommand{\tabref}[1]{Table~\ref{#1}}
\newcommand{\secref}[1]{\S~\ref{#1}}

\newcommand{\erg}{{\rm erg}}
\newcommand{\sr}{{\rm sr}}
\newcommand{\s}{{\rm s}}
\newcommand{\au}{{\rm AU}}
\newcommand{\Hz}{{\rm Hz}}
\newcommand{\eV}{{\rm eV}}
\newcommand{\cm}{{\rm cm}}
\newcommand{\pc}{{\rm pc}}
\newcommand{\msun}{{\rm M}_{\odot}}

\newcommand{\mstar}{m_*}

\newcommand{\rsun}{R_{\odot}}

\newcommand{\msunyr}{M_\odot\,{\rm yr}^{-1}}

\newcommand{\hii}{H{\sc ii}}
\newcommand{\yr}{{\rm yr} }

\newcommand{\Myr}{{\rm Myr} }
\newcommand{\Gyr}{{\rm Gyr} }

\newcommand{\dd}{{\rm d}}
\newcommand{\cc}{{\rm cm}^{-3}}
\newcommand{\K}{{\rm K}}

\defcitealias{Wise2019}{W19}

\usepackage{xparse,soul}

\title[SMS formation in a massive cloud with H$_2$ molecules]
{Radiative feedback for supermassive star formation in a massive cloud with H$_2$ molecules in an atomic-cooling halo}

\author[Y. Sakurai, Z. Haiman and K. Inayoshi]
{Yuya Sakurai$^{1,2,3}$\thanks{yuya-sakurai@g.ecc.u-tokyo.ac.jp}, 
Zolt\'an Haiman$^{4}$, Kohei Inayoshi$^{2}$
\\ \\
$^{1}$Kavli Institute for the Physics and Mathematics of the Universe (WPI), the University of Tokyo, Kashiwa, Chiba, 277-8582, Japan\\
$^{2}$Kavli Institute for Astronomy and Astrophysics, Peking University, Beijing 100871, China\\
$^{3}$School of Physics, Georgia Institute of Technology, Atlanta, GA 30332, USA\\
$^{4}$Department of Astronomy, Columbia University, New York, NY 10027, USA\\
}
\begin{document}

\date{Draft version \today}

\maketitle

\label{firstpage}

\voffset=-0.4in

%=============%
%==  Abstract  ===%
%=============%
\begin{abstract}
Recent three-dimensional cosmological simulations of protogalaxy
formation have suggested that supermassive stars (SMSs) can form in
gas clouds in which H$_2$ 
cooling is suppressed by dynamical heating prior to the activation of atomic cooling~\citep{Wise2019}, 
but they stopped short of the following
growth of a central protostar.  Here we examine whether accretion
on the protostellar core in this cloud is sufficiently rapid, in the
face of the radiation feedback, to produce a SMS.  We perform
one-dimensional radiation-hydrodynamical simulations of the hot
collapsing cloud with non-equilibrium chemical reactions directly
adopting the cloud properties from \citet{Wise2019} as an initial
condition.  We find that the stellar Lyman-Werner (LW) radiation 
from the SMS
dissociates H$_2$ in the inner regions of the gas flow, increasing gas
temperature and thermal pressure, and temporarily stopping the
accretion.  However, this negative feedback ceases when the
self-gravity and inward ram pressure force on larger scales push the
gas inward.  The central protostar is unable to expand an \hii~region
due to the high density, and grows to a mass of $\gtrsim10^5\,\msun$.
Our results suggests the successful formation of SMSs, and resulting
massive ($\sim10^5\,\msun$) remnant black holes in the clouds, 
but need to be confirmed in two- or three-dimensional
simulations.
\end{abstract}

%==============%
%==  Keywords  ===%
%==============%

%===============%
%==  Introduction  ==%
%===============%
\section{Introduction}
\label{sec:introduction}

%Observations of SMBHs
Observations of distant quasars over the past two decades have shown
that supermassive black holes (SMBHs) with masses $\gtrsim10^9\,\msun$
exist at redshift $z\gtrsim 6$
\citep[e.g.,][]{Fan+2001,Mortlock2011,Wu2015,Matsuoka2016,Banados2018,Reed2019,Onoue2019,Yang2020}.
The existence of these SMBHs means that a billion solar mass or more
can rapidly accumulate in a small region within $1\,\Gyr$.  The
physical mechanism(s) for when and how this occurs remain unknown.

%formation model Pop III, cluster formation%
One of the SMBH formation scenarios is that they grow from low-mass
black holes (BHs) with $\sim100-10^5\,\msun$ \citep[see e.g.][]{Inayoshi2019}.
There are several models for the formation of the small ``seed'' BHs.
Population~III (hereafter Pop~III) stars with masses of
$\gtrsim100\,\msun$ \citep{Hirano2014, Hirano2015} gravitationally
collapse at the end of their lives, leaving remnant BHs with masses of
$\sim100\,\msun$ \citep{Heger2003}.  More massive stars with
$\gtrsim1000\,\msun$ can form through runaway stellar collisions in
dense primordial star clusters
\citep{Sakurai2017,Boekholt2018,Reinoso2018,Sakurai2019, Tagawa2020}.
These stars gravitationally collapse to intermediate-mass BHs (IMBHs)
with similar masses.

%Direct collapse
It has been pointed out that the subsequent growth of seed BHs
through gas accretion can be suppressed by radiation feedback
(\citealt{Ciotti2001,Milosavljevic2009,Novak2011,Park2011}; but see
also \citealt{Inayoshi2016} for the possible solution with
super-Eddington BH accretion).  To alleviate the slow growth caused by
this suppression, one possibility is the so-called ``direct
collapse'', in which larger seed BHs with masses $\gtrsim 10^5\,\msun$
are produced \citep{Loeb1994,Oh2002,Bromm2003,Begelman2006}.  In this
model, a protostellar core forms in the center of a gas cloud
surrounded by a dark matter halo with a virial mass of
$\gtrsim10^7\,\msun$. These so-called ``atomic cooling'' haloes are
larger than the ``minihaloes'' ($\sim10^{5-6}\,\msun$) where the first
Pop III stars form.  The gas in the atomic-cooling halo cools 
mostly via HI
lines, and the central protostar can grow via rapid gas accretion into
a supermassive star (SMS) with a mass $\gtrsim10^5\,\msun$.  The SMS
gravitationally collapses to a BH with a similar mass due to general
relativistic instability \citep{Umeda2016, Woods2017, Haemmerle2018}.

%condition for the direct collapse
One of the conditions for SMS formation in atomic-cooling haloes is to
avoid ${\rm H_2}$--cooling induced fragmentation. This can be achieved
by irradiating the halo by an unusually strong external
far-ultraviolet Lyman-Werner (LW) radiation emitted from nearby
star-forming galaxies \citep{Omukai2001, Dijkstra2008,
  Regan2014,Sugimura2014,Inayoshi2014b,Becerra2015,Latif2016,Chon2016}.
A sufficiently intense LW radiation fully removes H$_2$ molecules
and suppresses H$_2$ cooling in some rare situations: the halo has a
nearby neighbouring halo with highly synchronised
star-formation~\citep{Visbal2014,Regan+2017}, so that the LW radiation
flux could exceed a critical 
flux~\citep{Wolcott-GreenHaiman2019}.
To form an atomic-cooling halo
the gas in the halo also needs
to remain extremely metal-poor. 
Moreover, the halo needs to be free from tidal force dispersion
\citep[e.g.][]{Chon2016}.
If the H$_2$ molecules are fully
dissociated, the only effective coolant is atomic hydrogen. The gas
temperature can not fall below $\sim8000\,\K$, resulting in elevated
sound speeds and higher accretion rates than in cooler gas in
minihaloes.

%Dynamical heating scenario
In addition to radiative processes, the dynamical effect of collapsing
gas into a massive DM halo as the halo is assembled is expected to
play a crucial role on the formation of SMSs, by suppressing ${\rm
  H_2}$--cooling~\citep{Fernandez+2014} and heating the halo gas via
compression and shocks~\citep{Yoshida+2003}, especially in haloes with
unusually rapid assembly histories.  Recently,
\citet[][hereafter \citetalias{Wise2019}]{Wise2019} have shown,
using three-dimensional cosmological hydrodynamical simulations, that
strong dynamical heating helps to keep the gas warm prior to the atomic-cooling stage, and may 
lead to the formation of SMSs with masses $\gtrsim10^5\,\msun$ 
in massive haloes with extremely rapid growth rates,
even if strong external LW radiation is absent, and H$_2$ molecules
are not fully dissociated.  Similarly, unusually large baryonic
streaming velocities can delay the collapse of gas into less massive
haloes and induces violent mergers of gaseous haloes into more massive
DM haloes \citep{Hirano2017,Schauer2017,Inayoshi2018}, inducing
dynamical heating and helping to produce the conditions required for
SMS formation.

%Radiation feedback can be problematic in the DH scenario
Although dynamical heating in unusually rapidly assembling haloes is
one of the promising mechanisms for the formation of SMSs, it remains
unclear whether SMSs do indeed form in these haloes. In particular,
radiation feedback from the growing protostar can stunt its growth,
when the accretion rate is lower than a critical value of 
$\approx
0.004-0.1\,\msunyr$ which is determined by equating the total luminosity of 
the star to the Eddington luminosity 
~\citep{Omukai2001b,Omukai2003,Hosokawa2013, Schleicher2013,Sakurai2015,Haemmerle2018}.
We hereafter adopt the conservatively high $0.04\,\msunyr$ for the critical rate, following \citet{Hosokawa2013}.
Above this critical rate, the rapid gas accumulation, as well as heat
input owing to rapid accretion prevent the stellar surface from
contracting via thermal emission on the Kelvin-Helmholtz (KH)
timescale. The rapidly accreting protostars evolve to `super-giants'
which have inflated radii of $\sim 100\,\au$.  Conversely, since the
KH timescale at the surface is $10^{3-4}\,\yr$, if the protostar grows
at rates lower than the critical value at the beginning of its
evolutionary stage the protostar cannot evolve into the supergiant
protostar.  The star contracts to a small radius, and develops a
correspondingly high effective temperature $\sim10^5\,\K$, emitting
copious amounts of ionising photons, which cause the radiation
feedback.  At their last resolved snapshots of the collapsing gas in
the three-dimensional simulations by \citetalias{Wise2019}, the accretion rates in the
innermost regions fall below this critical value (see the bottom right
panel of their Figure~4), leaving the fate of the protostar unclear.

%This work
In this study, we explore the evolution of gas inflows around the
growing protostar, and past the epoch simulated in \citetalias{Wise2019}, using
one-dimensional radiation hydrodynamical simulations. We adopt the
initial conditions for the cloud properties directly from \citetalias{Wise2019}, and
employ non-equilibrium chemical reactions.  We include the radiation
emitted by the growing protostar (as well as a tentative circumstellar
disc).  The main goal of this study is to assess whether the stellar
radiation suppresses the accretion rate below the critical value, or
if accretion remains sufficiently rapid to produce a SMS.

%structure of this paper
The rest of this paper is organised as follows.  In
\secref{sec:methods}, we describe the setup and details of the
radiation-hydrodynamical simulations.  In \secref{sec:results}, we
present the evolution of the gas clouds for cases with and without
stellar radiation.  In \secref{sec:discussion}, we discuss our
results, including their implications and some caveats.  Finally, in
\secref{sec:summary}, we summarise our main conclusions.

%===============%
%==  Methods  ==%
%===============%
\section{Methods}
\label{sec:methods}
\subsection{Hydrodynamical simulations}
\label{sec:Hydrodynamical simulations}

%ZEUS code
In order to explore gas inflows around growing protostars, we use the
hydrodynamical simulation code ZEUS~\citep{Stone1992}, including
multifrequency radiation transfer, photoionisation and heating, and a
primordial chemical network \citep{Inayoshi2016,Sakurai2016}.
Assuming spherical symmetry, we use ZEUS to solve the continuity
equation and the equation of motion in one dimension,
\begin{align}
  \pddt{\rho}{t}+\frac{1}{r^2}\pddt{}{r}(r^2\rho v)&=0, \\ \rho
  \left(\pddt{v}{t}+v\pddt{v}{r}\right)&=-\pddt{p}{r}-\rho\pddt{\Phi}{r}+f_{\rm
    rad},
\end{align}
where $t$ is time, $r$ is the radial coordinate, $\rho$ is the gas
density, $v$ is the velocity (defined to be negative for inflows),
$p=(\gamma-1)\rho e$ is the gas pressure, $\gamma=5/3$ is the
adiabatic index, $e$ is the specific internal energy of the gas,
$\Phi$ is the gravitational potential (including contributions from
both the growing protostar as a point mass, and from the self-gravity
of the gas) and $f_{\rm rad}$ is the radiation pressure force.  The
specific internal energy $e$ is determined by the energy equation
\begin{align}
\rho\left(\pddt{e}{t}+v\pddt{e}{r}\right)=-p\frac{1}{r^2}\pddt{}{r}(r^2
v) -\Lambda +\Gamma,
\end{align}
where $\Lambda$ and $\Gamma$ are the cooling and heating rates.
$\Lambda$ includes line cooling by H, H$_2$, H$_2^+$ and He,
recombination cooling of H$^+$ and He$^+$, free-free emission,
collisional ionisation cooling of H and He and H$_2$ dissociation
cooling \citep{Glover2007, Glover2008},
\begin{align}
\Lambda&=\Lambda_{\rm H}+\Lambda_{\rm H_2}+\Lambda_{\rm
  H_2^+}+\Lambda_{\rm He} \nonumber \\ &+\Lambda_{\rm
  H^+,rec}+\Lambda_{\rm He^+,rec}+\Lambda_{\rm ff}+\Lambda_{\rm H,col}
\nonumber \\ &+\Lambda_{\rm He,col}+\Lambda_{\rm
  H_2,dis}. \label{eq:cooling}
\end{align}
We omit the He cooling rate by the 2$^3$S metastable excitation state
which is proportional to $n_{\rm e}^2n_{\rm He^+}$ since in our
density regime this cooling rate can be invalid (see equation 14.19-20
in \citealt{Draine2011}).  For the heating rate $\Gamma$, we include
H$_2$ formation heating, photoionisation heating of H, He, He$^+$ and
H$_2$, H$^-$ photo-detachment heating, and H$_2$ photodissociation
heating \citep{Abel1997, Omukai2000}.

%chemical reactions
Our non-equilibrium chemistry incorporates the nine species H, H$^+$,
He, He$^+$, He$^{++}$, e$^-$, H$_2$, H$_2^+$ and H$^-$.  The chemical
reactions are taken mainly from Nos. 1-32 in Table A1 of
\citet{Glover2008}.  We adopt the case B recombination rates for
H$^+$, He$^+$ and He$^{++}$ because diffusive photons produced by
direct recombination to the ground states are immediately absorbed by
the surrounding gas.  We also consider photoionisation, H$^-$
photodetachment and H$_2$ photodissociation, with the rates adopted
from references listed in \tabref{tab:photoionisation}.  The evolution
of the number density of each species $i$ is governed by
\begin{align}
\pddt{n_i}{t}=C_i-D_in_i,
\end{align}
where $C_i$ and $D_i$ are creation and destruction terms of species
$i$ respectively.  The equation is solved using a semi-implicit method
updating each species in order \citep{Anninos1997, Whalen2006}.  The
order of the updates is H, H$^+$, He, He$^+$, He$^{++}$, H$^-$,
H$_2^+$, H$_2$ and e$^-$.  We set a calculation timestep as the
smallest among the Courant time, the cooling/heating time and the
chemical time.  The latter two timesteps are defined by
\begin{align}
t_{\rm cool} &= 0.1\frac{\rho e}{|\Lambda-\Gamma|}, \\ t_{\rm chem} &=
0.1\frac{y_{{\rm H}^+}+0.001(y_{\rm H}+y_{{\rm H}_2})}{\dot{y}_{{\rm
      H}^+}},
\end{align}
where $y_{{\rm H}}\equiv n_{{\rm H}}/n$, $y_{{\rm H}^+} \equiv n_{{\rm
    H}^+}/n$, $y_{{\rm H}_2}\equiv 2n_{{\rm H}_2}/n$ and $n$ is the
number density of hydrogen nuclei \citep{Whalen2006}.

\begin{table}
  \begin{center}
    \caption{H and He photoionisations, H$^-$ photo-detachment and
      H$_2$ photodissociation. }
    \begin{tabular}{ll} \hline
    Reactions & Ref.\\ \hline ${\rm H}~~~+~~\gamma ~~\rightarrow~~
    {\rm H}^+~~+~~{\rm e}^-$ & 1 \\ ${\rm He}~~+~~\gamma
    ~~\rightarrow~~ {\rm He}^+~+~~{\rm e}^-$ & 2 \\ ${\rm
      He}^++~~\gamma ~~\rightarrow~~ {\rm He}^{++}+~~{\rm e}^-$ & 3
    \\ ${\rm H}_2~~+~~\gamma ~~\rightarrow~~ {\rm H}_2^+~~+~~{\rm
      e}^-$ & 1 \\ ${\rm H}^-~~+~~\gamma ~~\rightarrow~~ {\rm
      H}~~~+~~{\rm e}^-$ & 4 \\ ${\rm H}_2^+~~+~~\gamma
    ~~\rightarrow~~ {\rm H}~~~+~~{\rm H}^+$ & 4 \\ ${\rm
      H}_2^+~~+~~\gamma ~~\rightarrow~~ 2{\rm H}^+~~+~~{\rm e}^-$ & 5
    \\ ${\rm H}_2~~+~~\gamma ~~\rightarrow~~ {\rm H}_2^* \rightarrow
       {\rm H}~~+~~{\rm H}$ & 5 \\ \hline
    \end{tabular}
    \begin{tablenotes}
    \small 1. \citet{Hui1997}, 2. \citet{Yan1998},
    3. \citet{Verner1996}, 4. \citet{Tegmark1997}, 5. \citet{Abel1997}
    \end{tablenotes}
    \label{tab:photoionisation}
  \end{center}
\end{table}

%radiative transfer
We solve the multi-frequency radiation transfer equation assuming a
steady-state radiation field because the photon crossing time ($\sim
\tau r/c$) is much shorter than the hydrodynamical timescale of gas
inflows.  Since ionised gas is optically thin to 
extreme- and far-ultraviolet (EUV and FUV)
radiation, we assume $F_\nu = cE_\nu$, where $F_\nu$ and $E_\nu$ are
the specific radiation flux and energy density. Then, if no other
radiation sources exist, the radiative transfer equation reduces to
\begin{align}
F_{k,\nu}=\left(\frac{r_{k-1}}{r_k}\right)^2 F_{k-1,\nu}
\exp\left[-(r_k-r_{k-1})\sum_i
  n_i\sigma_{i,\nu}\right], \label{eq:F_k}
\end{align}
where the subscript $k$ marks the radial cell and $\sigma_{i,\nu}$ is
the absorption cross section for species $i$ \citep{Whalen2006}.  We
do not treat diffuse EUV and FUV photons emerging by
recombination and radiative de-excitation in the radiative transfer equation of Eq. (\ref{eq:F_k}).
Instead of considering photoionisation by diffuse EUV photons, we here
adopt the on-the-spot approximation where the case B recombination
rate coefficient is used.  The flux of diffuse FUV radiation is
negligible compared to those from the central protostar and an
external LW background flux (see below).  We adopt the cross sections
from the references shown in \tabref{tab:photoionisation}, except for
LW radiation.

%LW radiation
For LW radiation, we replace the exponential factor in
Eq.~\eqref{eq:F_k} by a shielding factor that includes both
self-shielding of H$_2$, and shielding of H$_2$ by neutral H,
\begin{align}
  &f_{{\rm sh},k+1}(N_{\rm H_2},N_{\rm HI},T_,r)=\min(f_{{\rm
      sh,H_2},k+1}\times f_{{\rm sh,HI},k+1}, f_{{\rm sh},k})
  \\ &f_{\rm sh,H_2}=\frac{0.965}{(1+x/b_5)^{1.1}} \nonumber
  \\ &~~~~~~~~+\frac{0.035}{(1+x)^{0.5}}\exp\left[-8.5\times
    10^{-4}(1+x)^{0.5}\right] \\ &f_{\rm sh,HI}=(1+x_{\rm
    HI})^{-1.6}\exp(-0.15 x_{\rm HI})
\end{align}
where $f_{{\rm sh},0}=f_{{\rm sh,H_2},0}f_{{\rm sh,HI},0}$, $x\equiv
N_{\rm H_2}/(5\times10^{14}\,\cm^{-2})$, $b_5\equiv\sqrt{kT/m_{\rm
    p}}/(10^5\,\cm\,\s^{-1})$ and $x_{\rm HI}\equiv N_{\rm
  HI}/(2.85\times10^{23}\,\cm^{-2})$ \citep{Wolcott-Green2011}. To
obtain local estimates of the H$_2$ and HI column densities, we adopt
the `Sobolev-like' length defined as $L'_{\rm Sob}\equiv
\rho/|\dd\rho/\dd r|$ and use the relations $N_{\rm H_2}=n_{\rm
  H_2}L'_{\rm Sob}$ and $N_{\rm HI}=n_{\rm HI}L'_{\rm Sob}$.  We
include the LW radiation from the central protostar, as well as an
external LW background with specific intensity $J_{\rm
  LW}=3\times10^{-21} \erg\,\cm^{-2}\,\s^{-1}\,\Hz^{-1}\,\sr^{-1}$
(the same value as in \citetalias{Wise2019}).  Since the background radiation irradiates
the cloud from the outside in, whereas the stellar radiation
irradiates the cloud from the center, we define the self-shielding
factor outside-in for the background radiation, and inside out for the
stellar radiation.  Specifically, the self-shielding factor for the
background radiation is computed as $f_{{\rm sh},k-1}=\min(f_{{\rm
    sh,H_2},k-1}f_{{\rm sh,HI},k-1},f_{{\rm sh},k})$ with $f_{{\rm
    sh},k_{\rm max}}=f_{{\rm sh,H_2},k_{\rm max}}f_{{\rm sh,HI},k_{\rm
    max}}$.

%photon processes
The reaction rates $k_i$ and photo-heating rates $\Gamma_i$ for
photoionisation, H$^-$ photo-detachment and H$_2$ photodissociation
(\tabref{tab:photoionisation}) are computed using a photon-conserving
scheme \citep{Whalen2006} as
\begin{align}
k_i &=\int_{\nu_{{\rm th},i}}\frac{4\pi
  \hat{J}_\nu}{h\nu}\sigma_{i,\nu}\dd\nu, \\ \Gamma_i &= n_i
\int_{\nu_{{\rm th},i}}\frac{4\pi
  \hat{J}_\nu}{h\nu}\sigma_{i,\nu}E_{{\rm heat},i}\dd\nu,
\end{align}
where $\nu_{{\rm th},i}$ is the threshold frequency for species $i$,
$\hat{J}_\nu$ is the mean intensity (over solid angles) and $E_{{\rm
    heat},i}\equiv h(\nu-\nu_{{\rm th},i})$.  The photon conservation
method here means that the absorbed flux contributes to the estimation
of ionisation and heating rates, assuming that all the excess energy
is thermalized and deposited into the gas.  For the two-step H$_2$
dissociation by LW photons (often referred to as the ``Solomon
process''; \citealt{Field+1966}), we adopt the reaction rate
\begin{align}
k_{\rm LW}=1.1\times10^8\,\frac{F_{k,\nu}}{{\rm
    erg\,s^{-1}\,cm^{-2}\,Hz^{-1}}}~{\rm s^{-1}}
\label{eq:Solomon rate}
\end{align}
and the heating rate
\begin{align}
\Gamma_{\rm LW}=6.4\times10^{-13}n_{{\rm H}_2}k_{\rm LW}\,{\rm
  erg\,s^{-1}\,cm^{-3}}.
\end{align}
\citep{Abel1997}. The radiation pressure force is
\begin{align}
f_{\rm rad}=\frac{n_{\rm e}}{c}\int\sigma_{\rm
  es}F_\nu\dd\nu+\frac{\Gamma}{c},
\end{align}
where $\sigma_{\rm es}$ is the Thomson cross section and $\Gamma$ is
the total bound-free photoheating rate.

%radiation source model 
For the central radiation sources, we consider their contributions from both the
growing protostar and from a hypothetical circumstellar disc.  We
include the disc component, because the gas contracting in the cores
of the haloes simulated in \citetalias{Wise2019} has non-negligible angular momenta
(see their Extended Data Figure 4), suggesting that a circumstellar
disc may form.  Both sources are unresolved and located at the origin
$r=0$.

%star radiation source and disc radiation source
The stellar radiation flux at the innermost cell is
\begin{align}
F_{*,\nu}=\pi\left(\frac{R_*}{r_{\rm min}}\right)^2 B_\nu(T_{\rm
  eff}),
\end{align}
where $R_*$ is the stellar radius, $T_{\rm eff}$ is the effective
temperature, $r_{\rm min}$ is the radius of the innermost cell, and
$B_\nu$ is the Planck function.  The stellar radius and effective
temperature are calculated from a stellar evolution model described in
\secref{sec:SE}.  We adopt a standard disc model and a multicolor
blackbody spectrum \citep[e.g.][]{Kato2008}, 
which is well approximated by a $\nu^{1/3}$
power-law in the UV range of interest, for computing the disc
radiation flux
\begin{align}
F_{{\rm disc},\nu} &=\frac{1}{6\pi r_{\rm min}^2 [(\nu_*/\nu_{\rm
      min})^{4/3}-1]\nu_{\rm min}}\frac{GM_* \dot{M}}{R_*}
\left(\frac{\nu}{\nu_{\rm min}}\right)^{1/3} \nonumber
\\ &~~~~~~~~~~~~~~~~~~~~~~~~~~~~~~~~~~~~~~~(\nu_{\rm min}\le \nu \le
\nu_*),
\end{align}
where
\begin{align}
\nu_* &=3.14\times 10^{15}\,\Hz \nonumber
\\ &\times\left(\frac{M_*}{1\,\msun}\right)^{1/4}
\left(\frac{\dot{M}}{10^{-2}\,\msunyr}\right)^{1/4}
\left(\frac{R_*}{1\,\rsun}\right)^{-3/4}.
\end{align}
Note that the cutoff frequency $\nu_{\rm cut}\equiv \nu_*$, which
corresponds to the frequency of the maximum flux of the optically
thick disc, always remains below the maximum frequency $\nu_{\rm
  max}\simeq 2.85\times10^{16}\,\Hz$ in our simulations.  Since
$h\nu_*$ is always below $\sim5\,\eV$, less than the EUV and FUV
energies, the disc radiation has only a relatively minor effect on the
dynamics of the flow.  The total flux is $F_{{\rm
    in},\nu}=F_{*,\nu}+F_{{\rm disc},\nu}$.

%grid generation
We adopt a spherically symmetric, logarithmically spaced grid in the
simulations.  The $k^{\rm th}$ cell of the grid is located at
$r_k=r_{\rm min}+(r_{\rm max}-r_{\rm min})
(\epsilon^{k-1}-1)/(\epsilon^N-1)$, where $N$ is the number of cells,
$r_{\rm max}$ is the radius of the outermost cell and $\epsilon$
($\Delta r_{\rm k+1}/\Delta r_{\rm k}$) is the ratio between the radii
of consecutive cells.  The adopted grid parameters are summarised in
\tabref{tab:grid} for convenience.  The innermost cell radius $r_{\rm
  min}$ (i.e. the inner boundary of the simulation) is chosen so that
it is comparable to the star's gravitational radius $R_{\rm
  B}=2GM_*/c_{\rm s,\infty}^2$ at the initial time, which is $\sim
8.2\times 10^{15}\,\cm$ assuming $T_\infty=300\,\K$ and
$M_*=2\,\msun$. 
The radius of the innermost cell is always
larger than the stellar radius of a highly accreting protostar with a
bloated envelope with $r \lesssim 2\times 10^{15}$ cm for $M_*
\lesssim 10^5 \msun$ \citep{Hosokawa2013}.
We use outflow boundary conditions at both the inner
and the outer boundary: gas is allowed to pass from simulation
regions to outsides but is not allowed to flow in. 
With these boundary conditions masses in the simulation domain
continuously decrease. In this case the accretion rate is artificially
decreased when the stellar masses, on to which most outflowing gas is
accreted, become comparable to the domain masses, and the inflowing 
gas is more prone to radiation feedback. We still adopt these 
conservative boundary conditions.

\begin{table}
  \begin{center}
    \caption{Grid parameters for our hydrodynamical simulations.}
    \begin{tabular}{ll} \hline
    $N$ & 600 \\ $r_{\rm min}~(\cm)$ & $10^{16}$ \\ $r_{\rm
        max}~(\cm)$ & $10^{20}$ \\ $\epsilon$ & 1.008 \\ $\nu_{\rm
        min}~(\Hz)$ & $10^{13}$ \\ $\nu_{\rm max}~(\Hz)$ & $2.85\times
      10^{16}$ \\ $N_{\nu}$ & $50$ \\ \hline
    \end{tabular}
    \label{tab:grid}
  \end{center}
\end{table}

%frequency bin
We adopt a frequency grid which allows us 
to follow the relevant
radiative processes (\tabref{tab:photoionisation}).  The frequency
range is $10^{13}\,\Hz<\nu<2.85\times10^{16}\,\Hz$ or
$0.04\,\eV<h\nu<118\,\eV$.  The number of frequency bins is
$N_{\nu}=50$.  We designed the grid layout to decrease the number of
frequency bins making computation time shorter: we choose a fine
frequency mesh at energies moderately larger than threshold energy of
each of the reactions in \tabref{tab:photoionisation} and space the
bins more sparsely at other photon energies.

%Initial conditions
The initial conditions of our simulations are taken from the
spherically-averaged gas cloud profiles of the LWH model in \citetalias{Wise2019} (see
dashed curves in their Fig. 4).  These include the gas density,
velocity, temperature, H$_2$ fraction and electron
fraction.  In their simulations they also show
results for the the cloud ``MMH'' which is their most massive halo.
The results for LWH and MMH are qualitatively similar in our
simulations so in the following we will focus on the results of the
LWH cloud for simplicity.  The initial profiles we adopted are shown
by the black curves in \figref{fig:wrad} below.  The initial H$^+$
fraction is set to that of electron because of the charge neutrality
(note that helium is neutral at the initial condition).  The ratio of
the number density of hydrogen nuclei to that of helium nuclei is
$0.0833$.  Helium is assumed to be initially all neutral.

%assumption of spherical symmetry
Although in our simulations we assume a spherically symmetric gas
distribution, in the 3-D simulation of \citetalias{Wise2019} the gas distribution is not
spherically symmetric.  We discuss the possible impact of the
spherical assumption on our results in \secref{sec:spherical
  assumption} below.

\begin{table*}
  \begin{center}
    \caption{The table shows fits to the stellar radii and effective
      temperatures of ZAMS stars as a function of stellar mass.  We
      use the data from \citet{Marigo2001} for $M_*\le 100\,\msun$ and
      from \citet{Bromm2001} for $M_*\ge300\,\msun$.}
    \begin{tabular}{llllllll} \hline
    $M_* ~(\msun)$           &  2                & 10      & 30      & 50       & 100     & 300   & 1000 \\ \hline
    $\log R_* ~(\rsun)$       &   -8.93e-4   & 0.139 & 0.323 &  0.451 &  0.627 & 0.959 & 1.20 \\
    $\log T_{\rm eff} ~(\K)$ &   4.14         & 4.65   & 4.87    & 4.93    & 4.98   & 5.05   & 5.07  \\ \hline
    \end{tabular}
    \label{tab:fitting SE1}
  \end{center}
\end{table*}

\begin{table*}
  \begin{center}
    \caption{The table shows fits to the stellar radii and effective temperatures of protostars which grow through constant accretion rates of $\dot{M}=10^{-1}\,\msunyr$.}
    \begin{tabular}{lllllllll} \hline
    $M_* ~(\msun)$            &   2       & 10      & 20    & 27     &  100    & 1.7e4  \\ \hline
    $\log R_* ~(\rsun)$       &   2.30   & 2.26   & 2.37 & 2.78  &  3.32  & 4.34 \\
    $\log T_{\rm eff} ~(\K)$ &   3.65   & 3.69   & 3.70 & 3.68  &  3.68  & 3.80  \\ \hline
    \end{tabular}
    \label{tab:fitting SE2}
  \end{center}
\end{table*}

\subsection{Stellar evolution}
\label{sec:SE}

%protostellar model
The growth of the central protostar during each time-step $\Delta t$
is calculated simply from $\Delta M_*=\dot{M}\Delta t$, using the mass
flux at the innermost cell as the accretion rate $\dot{M}$ onto the
protostar.  The protostellar evolution is computed by fitting stellar
evolution data.  Depending on the accretion rate, the evolution of a
rapidly growing protostar is divided into two phases: if the accretion
rate is lower than $\dot{M}_{\rm crit} (\equiv 0.04\,\msunyr)$, the
star is in a compact zero-age main sequence (ZAMS) phase, and
otherwise it is in a bloating phase \citep{Hosokawa2013}.  Even if the
accretion rate drops below $\dot{M}_{\rm crit}$, the star may be still
in the bloating phase for several thousand years \citep{Sakurai2015}.
However, we do not model this sustained bloating phase, and instead
assume the star is in the ZAMS phase whenever
$\dot{M}\lesssim\dot{M}_{\rm crit}$.  As seen in \S \ref{sec:results},
this treatment makes our conclusion conservative, because the EUV
luminosity from the compact ZAMS model with the same mass is
substantially higher than that in the bloating phase with a lower
effective temperature of $T_{\rm eff}\sim 5000$ K.
Specifically, for $\dot{M}<\dot{M}_{\rm crit}$ we use the data for
ZAMS stellar evolution from \citet{Marigo2001} and \citet{Bromm2001},
as summarised in \tabref{tab:fitting SE1}.  For $\dot{M}\ge
\dot{M}_{\rm crit}$, we use the model data of a super-giant protostar
growing at a constant mass accretion rate of
$\dot{M}=10^{-1}\,\msunyr$.  The data of the stellar radii and
effective temperature are generated by using a stellar evolution code
STELLAR that was originally developed in \citet{Yorke2008} and used in
\citet{Sakurai2015} (see \tabref{tab:fitting SE2}).  We note that the
evolution of a highly accreting protostar hardly depends on the
detailed time-evolution of the mass inflow rate as long as
$\dot{M}\gtrsim0.04\,\msunyr$ is satisfied.

We set the initial stellar mass to $M_*=2\,\msun$, which is chosen so
that the dynamical timescale at the inner-most cell ($r_{\rm
  min}=R_{\rm B}\propto M_*$) is not too short to follow gas dynamics
over a wide range of spacial scales.  The choice of a smaller initial
mass would not make the result qualitatively different because
radiative feedback does not affect the mass accretion rate until the
star grows to $\sim 4~\msun$ as seen in \S \ref{sec:results}.  For
stellar radii and temperatures in-between the masses or oustide the
mass range in Tables \ref{tab:fitting SE1} and \ref{tab:fitting SE2},
we interpolate/extrapolate linearly in the logarithmic quantities.

%===============%
%==  Results  ==%
%===============%
\section{Results}
\label{sec:results}

\subsection{Simulations with and without radiation}
\label{sec:sims w/ and w/o rad}

\begin{figure*}
    \centering
    \includegraphics[width=1.0\textwidth]{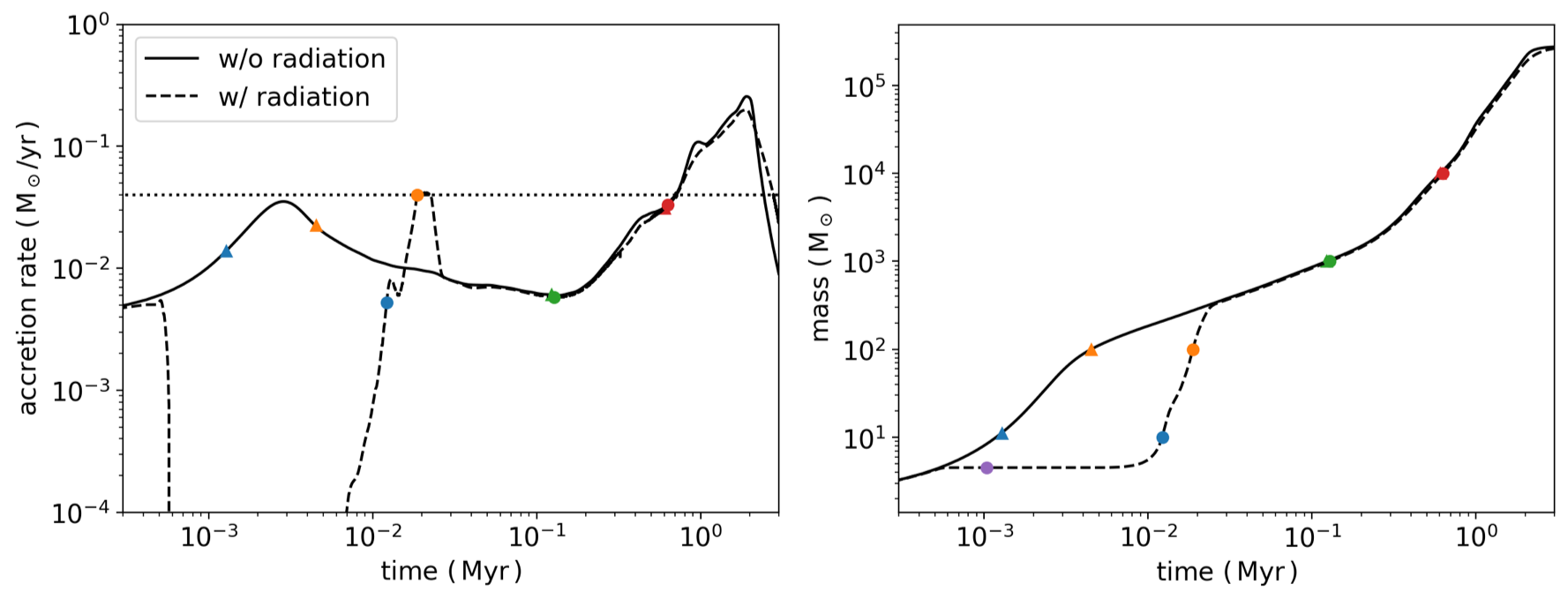}
    \caption{Left panel: evolution of the accretion rate.  Right
      panel: evolution of the mass of the central protostar.  In both
      panels, the solid curves show results without radiation, and the
      dashed curves show results with radiation included.  The
      horizontal dotted line indicates the critical accretion rate
      $0.04\,\msunyr$. The circles and triangles indicate the points
      where the stellar masses become $4, 10, 100, 10^3$ and
      $10^4\,\msun$ for the simulations with and without radiation
      respectively.}
    \label{fig:mass_acc}
\end{figure*}

%figure
In the left panel of \figref{fig:mass_acc}, we show the evolution of
the accretion rates at the inner boundary with time.  
The solid and dashed curves indicate
simulations without radiation and with radiation respectively.  The
horizontal dotted line shows the critical accretion rate
$\dot{M}=0.04\,\msunyr$ (see \secref{sec:introduction} and \secref{sec:SE}).  In the right
panel, we show the time evolution of the growing proto-stellar masses.

\begin{figure*}
    \centering
    \includegraphics[width=1\textwidth]{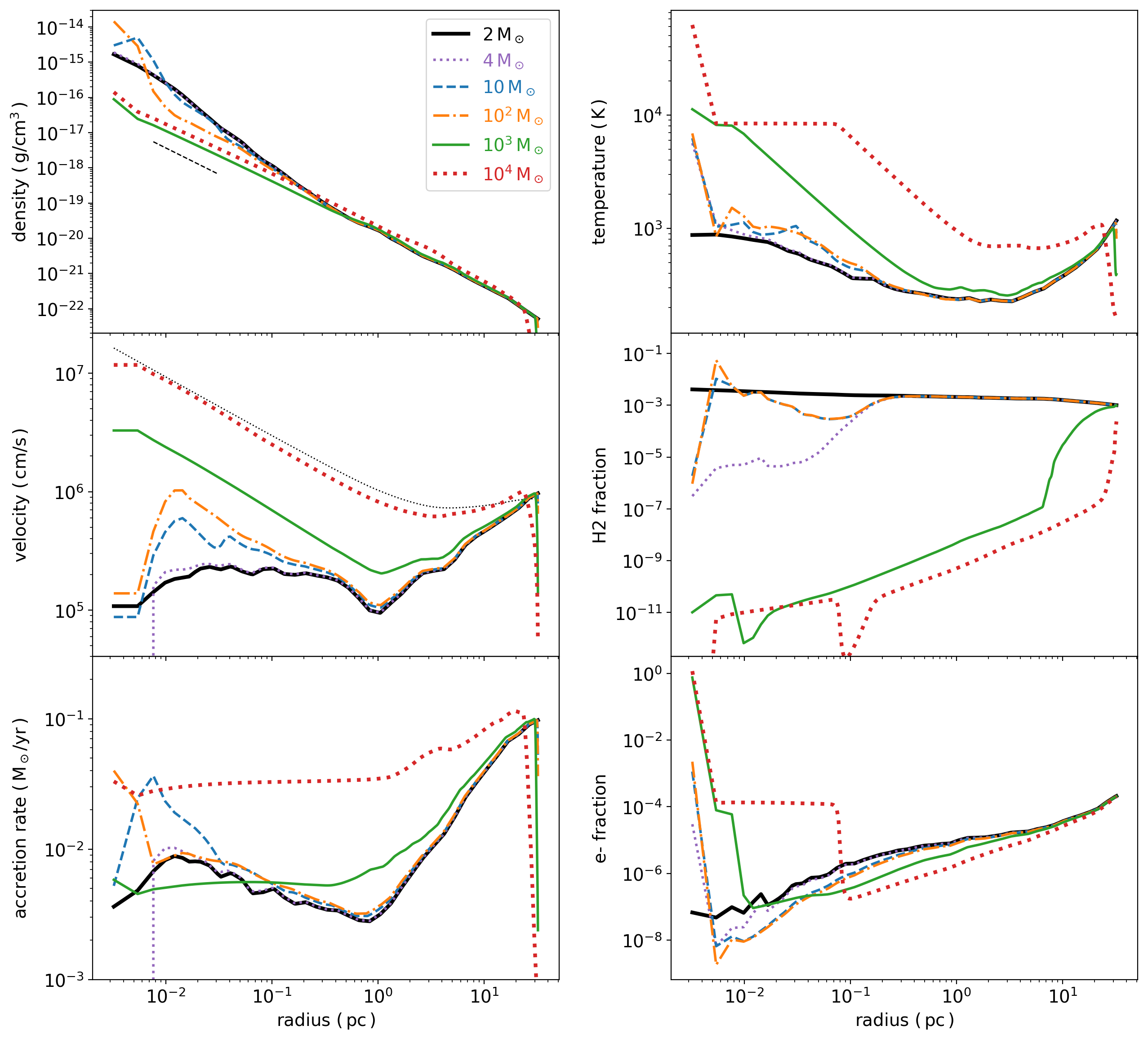}
    \caption{Radial profiles of density (top left), temperature (top
      right), velocity (middle left), H$_2$ fraction (middle right),
      accretion rates (bottom left) and electron fraction (bottom
      right) for the simulation with the radiation from the central
      source included.  We show snapshots of the profiles when the
      stellar masses are $2, 4, 10, 100, 1000$ and $10^4\,\msun$.  The
      thin dashed line in the density panel indicates the 
      slope $\propto
      r^{-1.5}$ expected for steady accretion via free-fall.  The
      thin dotted line in the velocity panel indicates the free-fall
      velocity at $M_*=10^4\,\msun$, $|v_{\rm ff}|=\sqrt{2G(M_*+M_{\rm
          enc})/r}$.  Note that for $M_*\gtrsim 10^3\,\msun$ the
      velocity of the flow is highly supersonic near the center.}
    \label{fig:wrad}
\end{figure*}

In the no-radiation case, the accretion rate remains above $\gtrsim
0.005\,\msunyr$ and the stellar mass grows monotonically without
interruptions.  In contrast, when stellar radiation is included, the
radiation stops the accretion onto the protostar temporarily for
$\sim10^4\,\yr$.  The growth of the stellar mass is halted at
$\sim4\,\msun$.  At $t\gtrsim10^4\,\yr$ accretion resumes, the rate
eventually increases to $\gtrsim10^{-2}\msunyr$, and the protostar
rapidly increases its mass by $\sim 2$ orders of magnitude in $\sim
{\rm a~few}\times 10^4$ yrs.  During this rapid accretion episode, the
accretion rate reaches the critical rate $\dot{M}_{\rm crit}$ at $t
\sim 3\times 10^4$ yr, but decreases after the peak 
because the density is decreased after gas of a few hundreds 
of solar masses accretes on to the protostar 
(the top left panel of \figref{fig:wrad}).
The drop of the accretion rate at $t\gtrsim 2\,\Myr$ are due to the
depletion of gas from the simulation domain. The drop is hardly affected
by the stellar radiation.

\begin{figure}
    \centering
    \includegraphics[width=0.5\textwidth]{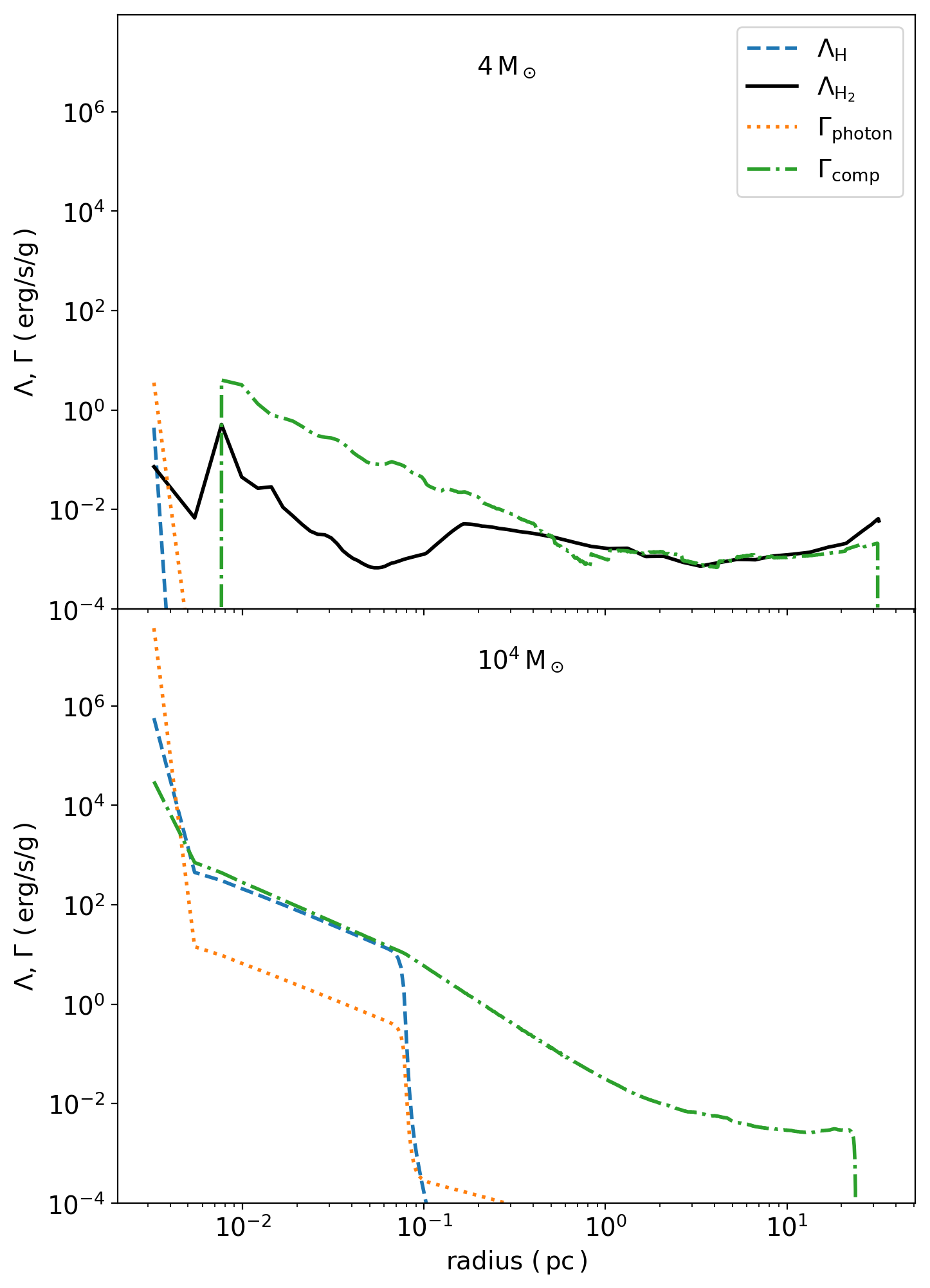}
    \caption{Radial profiles of cooling and heating rates for the
      simulation with radiation when $M_*\sim4$ (top) and
      $10^4\,\msun$ (bottom).  We show the H$_2$ cooling rate (black solid),
      the compressional heating rate (green dot-dashed), the atomic
      hydrogen line cooling rate (blue dashed) and the photoheating rate
      (orange dotted).  In the bottom panel, the atomic-cooling region
      appears at $\lesssim 0.1\,\pc$ where the gas is
      slightly ionised by the collisional ionisation of neutral
      hydrogen (see the bottom right panel of \figref{fig:wrad}). In
      the innermost region when $M_*\sim10^4\,\msun$, the photoheating
      rate is balanced by the ionised hydrogen recombination cooling
      rate plus the free-free emission cooling rate (not shown in the
      figure for clarity).}
    \label{fig:wrad_cool}
\end{figure}

%w/ radiation
In \figref{fig:wrad}, for the simulation with the radiation from the
central source included, we show snapshots of the radial profiles of
the gas density (top left), temperature (top right), velocity (middle
left), H$_2$ fraction (middle right), accretion rates (bottom left)
and electron fraction (bottom right) when the stellar masses are
$M_*=2, 4, 10, 100, 1000$ and $10^4\,\msun$.  The initial density
profile (black line in the top left panel) has a slope $\rho\propto
r^{-1.5}$ at $r\gtrsim 1\,\pc$, which is shallower than the
power-law $r^{-2}$ for isothermal collapse.  The shallower slope is
due to sheet-like structures seen in the 3-D simulation of \citetalias{Wise2019}.
We also show the profiles of the cooling/heating rates at the first
and last snapshot in \figref{fig:wrad_cool}, where the different
colors indicate the H$_2$ line cooling rate (black), compressional
heating rate (green), atomic hydrogen line cooling rate (blue) and
photoheating rate (orange).

%explanation for the fig
The gas inflow is stopped at $M_*\sim4\,\msun$ from $t\sim600\,\yr$ to
$\sim7000\,\yr$: the infall velocity and the accretion rate become
zero at $r\lesssim0.01\,\pc$ (purple curves in
\figref{fig:wrad}).  At $t\gtrsim10^4\,\yr$, the gas inflow resumes
and the stellar mass increases to $M_*\gtrsim10\,\msun$.  In the
accretion phase of $M_*\gtrsim 10^3\,\msun$, 
the slope of the density profile in the inner regions
$r\lesssim 0.3\,\pc$ gradually evolves to $\rho \propto r^{-1.5}$.
This change in slope occurs because the stellar gravitational radius
$R_{\rm B}$ increases to $\gtrsim 0.3\,\pc$, making gas free-fall
in the star's point-like gravitational potential (left middle panel).
The temperature reaches $\sim8000\,\K$ for $M_*\gtrsim10^3\,\msun$ in
the inner region, where atomic-hydrogen cooling becomes effective (see
blue curve in the bottom panel of \figref{fig:wrad_cool}).  Once the
core becomes hot enough to collisionally dissociate H$_2$ ($T\gtrsim
3000$ K) and the H$_2$ column density drops within the core, LW
radiation produced from the central star propagates out and
effectively dissociates H$_2$ in the outer region 
($r\lesssim 5\,\pc$).  Within the stellar influence radius
($r<R_{\rm B}\sim 1\,\pc$), the temperature increases inward due to
compressional heating (bottom panel of \figref{fig:wrad_cool}) from
the equilibrium temperature of H$_2$ cooling $\sim200\,\K$.

As the stellar mass reaches $\sim 10^4~\msun$, where the accretion
rate is still below the critical value, a fully-ionised ($T>10^4$ K;
$r\lesssim 0.005\,\pc$) and partially ionised region ($T\simeq
8000$ K; $0.005~\pc\lesssim r \lesssim 0.07\,\pc$) form.  While in the partially ionised region, the
electron fraction is determined by the balance between collisional
ionisation of neutral hydrogen (by electrons) and radiative
recombination of hydrogen, the inner-most hot region is created
because of photo-ionising radiation from the $\sim 10^4~\msun$ star
with $T_{\rm eff}\sim 10^5$ K.  Despite the strong stellar EUV
radiation, the gas is not fully ionised to form a large \hii~region.
This is because the gas density is so high ($n\simeq 10^8~\cc$ at
$r\simeq r_{\rm min}$) that the hydrogen recombination rate is faster
than the ionisation rate, and the \hii~region is unable to propagate
away from the stellar surface.
The H$_2$ fraction still remains as low as $10^{-11}$-$10^{-7}$ because of H$_2$ collisional dissociation at $r<0.07\,\pc$ and 
LW photodissociation at $r>0.07\,\pc$.

\begin{figure}
    \centering
    \includegraphics[width=0.5\textwidth]{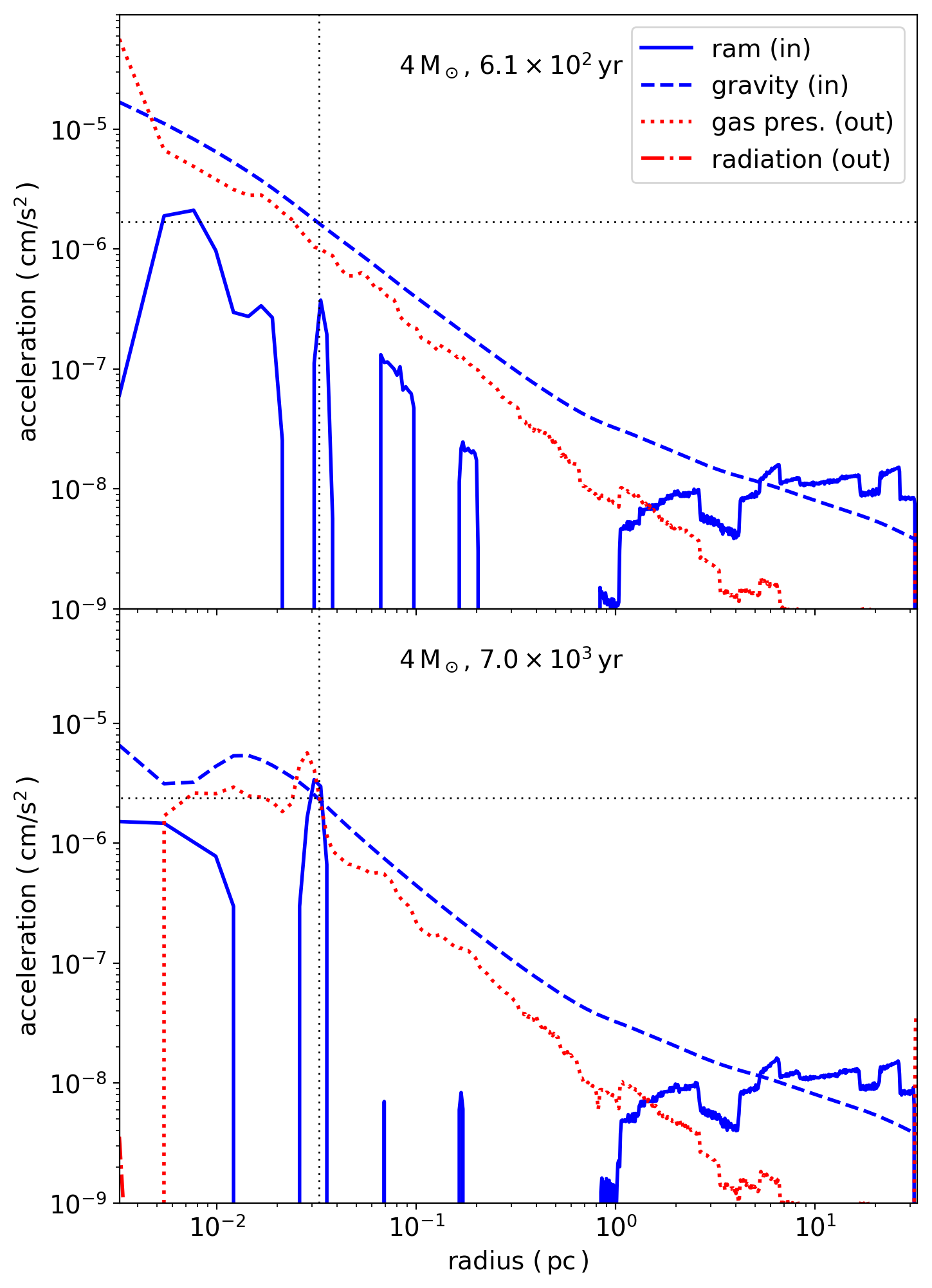}
    \caption{Radial profiles of forces for the simulation with a
      radiation source.  Two snapshots are shown, at $t\sim600$ and
      $7000\,\yr$, when the accretion first stops at
      $M_*\sim4\,\msun$, and just before it begins to resume, in the
      top and bottom panels, respectively.  We show the inward ram
      pressure force (solid), the stellar gravity plus gas
      self-gravity (dashed), the outward gas pressure gradient force
      (dotted) and the radiation pressure force (dot-dashed).  Inward
      forces are shown in blue and outward forces are shown in red.
      The dotted black vertical lines mark $r=10^{17}\,\cm\sim 0.03\,\pc$ and the
      dotted black horizontal lines mark the values of the
      gravitational acceleration at this radius.  The gravity
      increases from $1.7\times10^{-6}\,{\rm cm\,s^{-2}}$ to
      $2.4\times10^{-6}\,{\rm cm\,s^{-2}}$ between $t\sim600$ and
      $7000\,\yr$.  }
    \label{fig:force}
\end{figure}

%temporal feedback
We examine the reason why the gas inflow is temporarily stopped and
then resumes.  We show the cooling/heating rates at $t\sim10^3\,\yr$
and $M_*\sim4\,\msun$ when the gas inflow stops in the top panel of
\figref{fig:wrad_cool}.  The H$_2$ cooling rate is suppressed for
$r\lesssim 0.1\,\pc$ where the H$_2$ molecules are dissociated by
LW radiation (see the middle right panel of \figref{fig:wrad}).  
For $r \lesssim 0.04\,\pc$ photoheating and compressional heating are
effective, and HI and H$_2$ coolings are inefficient,
the temperature increases inward for $r\lesssim 0.01\,\pc$.
Since the sonic point moves inward with this increase of the
temperature, the outward gas pressure gradient force overcomes the
inward gravitational force on the gas, and decelerates the infalling
gas. This is seen directly in \figref{fig:force}, where the top panel
shows the radial profiles of the pressure and gravitational forces
at 610 yr, at the time when the accretion first stops, and reveals
that the outward pressure gradient force in the inner region becomes
dominant.

%accretion restarts
The accretion rate finally resumes at $t\gtrsim10^4\,\yr$,
  because the self-gravity of the gas builds up as the outer shells
  fall in and accumulate.  The over-pressurised region moves steadily
  outward from the inner core, until it reaches $r\sim 0.03\,\pc$ at
  $t\sim 10^4$ yr.  At the same time, the gas accumulating due to
  infall from larger radii increases both the inward gravitational and
  ram pressure forces.  For example at $r=10^{17}\,\cm\sim 0.03\,\pc$ the gravity
  increases from $1.7\times10^{-6}\,{\rm cm\,s^{-2}}$ at
  $t\sim600\,\yr$, just after the accretion stops, to
  $2.4\times10^{-6}\,{\rm cm\,s^{-2}}$ at $t\sim7000\,\yr$, just
  before the accretion recovers (upper vs. lower panel in
  \figref{fig:force}).  At $t\approx 1.4\times 10^4$ yr (not shown in
  \figref{fig:force}) the outward pressure force becomes
  subdominant at all radii, allowing accretion to resume.

\begin{figure*}
    \centering
    \includegraphics[width=1\textwidth]{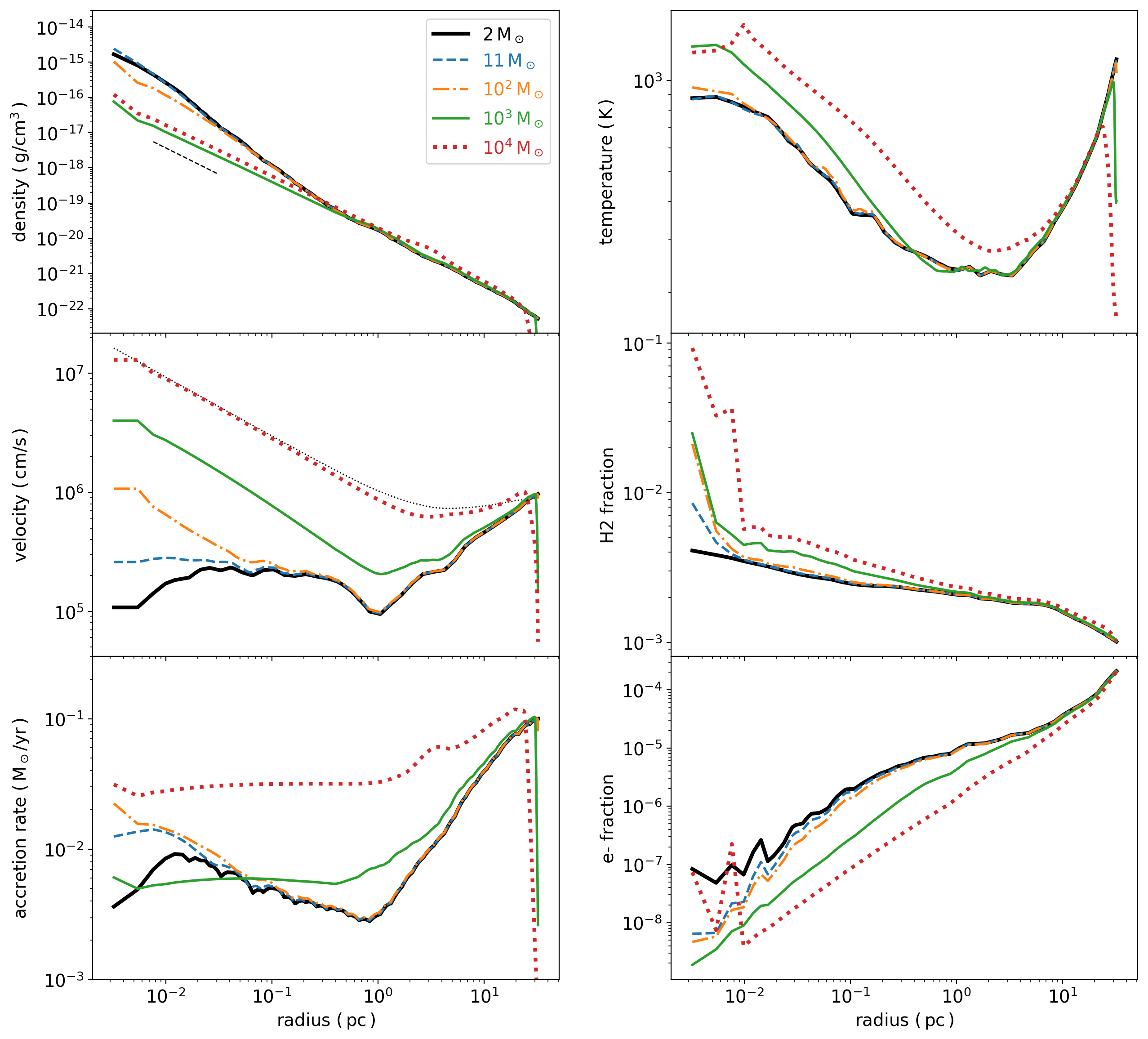}
    \caption{The same as \figref{fig:wrad} but for the simulation
      without radiation.  We show the profiles when the stellar mass
      is $2, 11, 100, 1000$ and $10^4\,\msun$.  }
    \label{fig:worad}
\end{figure*}

\begin{figure}
    \centering
    \includegraphics[width=0.5\textwidth]{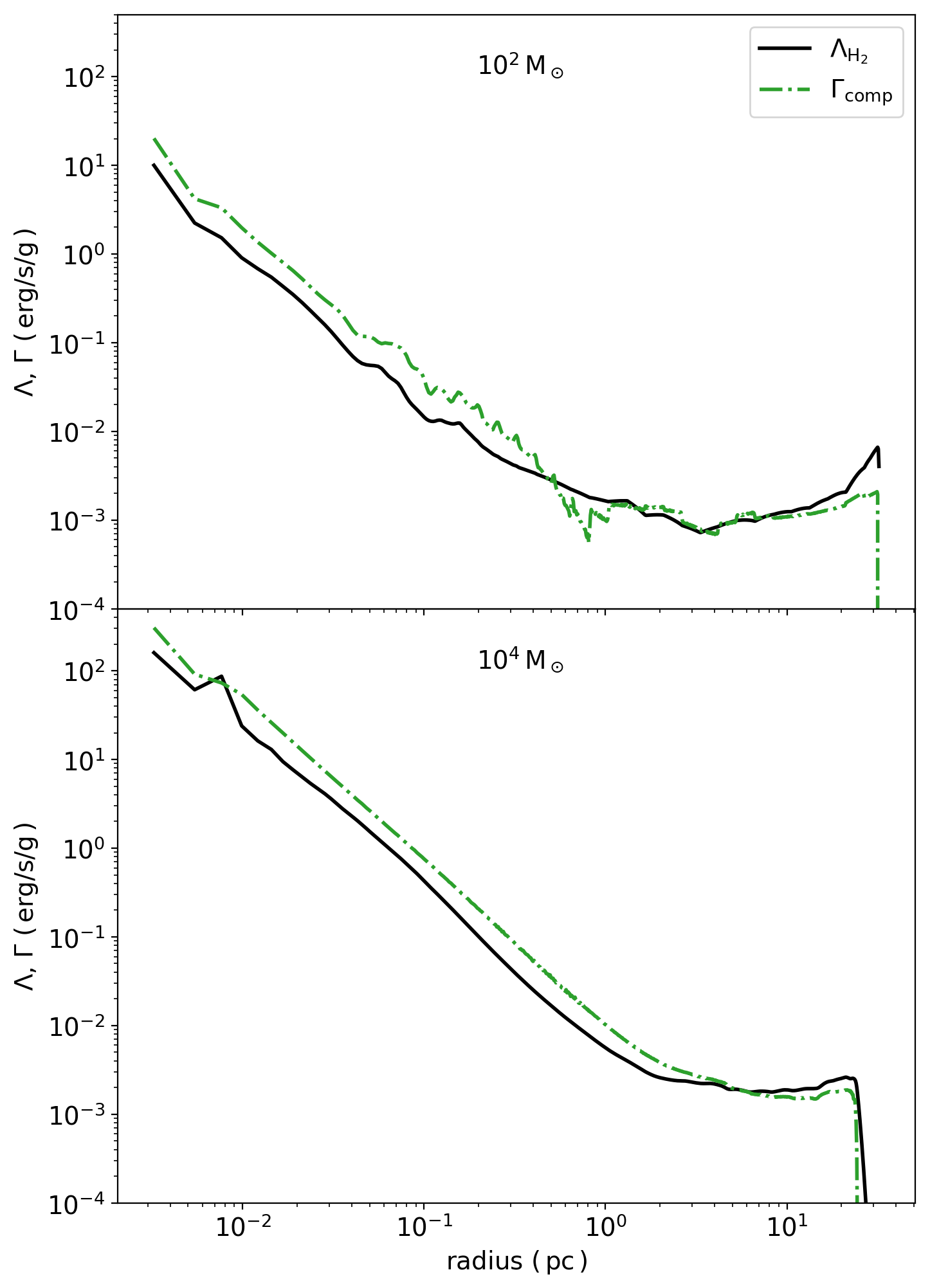}
    \caption{The same as \figref{fig:wrad_cool} but for the
      simulation without radiation.  We show the snapshots when the
      stellar mass is $M_*=100\,\msun$ (top panel) and
      $10^4\,\msun$ (bottom panel).  The total cooling rate is
      comparable to the H$_2$ cooling rate.  }
    \label{fig:worad_cool}
\end{figure}

%w/o radiation
For comparison, in \figref{fig:worad} we show radial profiles of the
density, temperature, velocity, H$_2$ fraction, accretion rates and
electron fraction for the no-radiation case, when the stellar masses
are $M_*=2, 11, 100, 1000$ and $10^4\,\msun$.  Likewise, in
\figref{fig:worad_cool}, we show profiles of cooling/heating rates
when the stellar masses are $M_*=100\,\msun$ and $10^4\,\msun$.  As
time elapses and the stellar mass grows to $M_*\gtrsim10^3\,\msun$,
the slope of the density profile in the inner regions
$r\lesssim1\,\pc$ gradually evolves from the isothermal one
$\rho \propto r^{-2}$ to the free-fall one $\rho \propto r^{-1.5}$ as
seen in the case with radiation field.  Because of the lack of stellar
radiation feedback, the inflow velocity is accelerated to the
free-fall value monotonically at all radii, and the accretion rate
($\dot{M}\propto r^2\rho |v|$) tracks its evolution without
suppression.  The gas temperature gradually increases inward but is
saturated once it reaches $\sim 2\times 10^3$ K, because the
compressional heating rate is balanced with the H$_2$-line cooling
rate in the case without stellar radiation (see
Fig. \ref{fig:worad_cool}).  In the central core ($n\gtrsim
10^8~\cc$), the H$_2$ fraction rises rapidly through the three-body
reaction ($3{\rm H}\rightarrow {\rm H}_2 +{\rm H}$) as seen in
pristine star forming clouds \citep{Yoshida2006,Turk2011}.

%===============%
%==  Discussion  ==%
%===============%
\section{Discussion}
\label{sec:discussion}
\subsection{Impact of the parameters}

The main result of our simulations is that formation of SMSs can take
place via rapid mass accretion because an \hii~region is unable to
propagate to large radii and hinder the inflow.  We expect that this
result may be changed by differences in the simulation setup and the
initial conditions.  Specifically, if the cloud had a lower initial
density, the accretion rate would be smaller and the SMS formation
could be inhibited if the \hii~region could expand.  Also, a stronger
LW radiation may help gas heat more, and hinder the gas infall.

\begin{figure}
    \centering
    \includegraphics[width=0.5\textwidth]{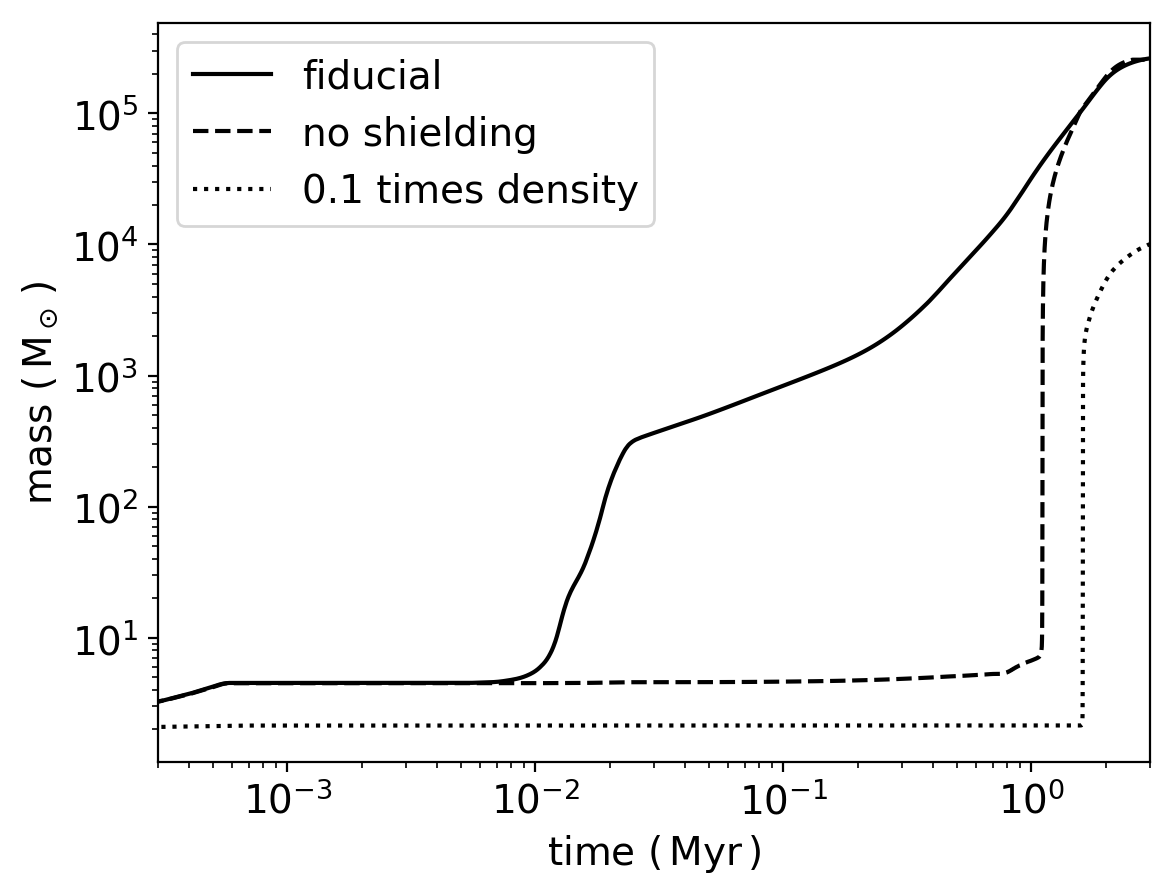}
    \caption{Evolution of the stellar mass. Solid curve: the fiducial
      simulation with radiation included. Dashed: the same simulation
      but with LW self-shielding ignored. Dotted: the fiducial
      simulation with the initial density profile reduced by a factor
      of ten.  }
    \label{fig:mass_acc2}
\end{figure}

To explore how these effects would impact our conclusions (i.e.
whether an SMS finally forms), we performed two variants of our
fiducial simulation.  First, in \figref{fig:mass_acc2}, we show the
evolution of the stellar mass for a simulation in which radiation is
included, but the H$_2$ self-shielding against LW radiation is ignored
(dashed curve).  In this case, the recovery of the accretion rate is
delayed from $\sim10^4\,\yr$ to $\gtrsim 1\,\Myr$.  The accretion rate
then increases to $\gtrsim1\,\msunyr$ and the stellar mass rapidly
increases from $\sim4\,\msun$ to $\gtrsim10^5\,\msun$ within
$\sim1\,\Myr$.  In this no-shielding model, the radiation feedback
process is similar to the fiducial model: the LW radiation dissociates
H$_2$ molecules, H$_2$ cooling becomes inefficient, the
gas temperature increases and the outward gas pressure gradient force
overwhelms the inward gravitational force.  The longer pause in the
stellar growth than in the fiducial model is due to more efficient
H$_2$ dissociation by the stronger (unshielded) LW radiation.

We next show the evolution of the stellar mass for a simulation in
which the initial density profile was assumed to be 10 times lower
than in the fiducial case (dotted curve in \figref{fig:mass_acc2}).
In this case, gas accretion is suppressed for $t\lesssim1.5\,\Myr$,
because the inner region of the cloud is initially gravitationally
stable due to the lower density.  After $t\gtrsim1.5\,\Myr$, the
accretion rate increases and reaches $\gtrsim0.1\,\msunyr$, because
gas from large scales falls inward and gravitational instability
develops.  The stellar mass increases from $\sim2\,\msun$ to
$\gtrsim10^4\,\msun$ until $t\sim3\,\Myr$.

We conclude that the SMS formation is viable if the density is larger
than 0.1 times the density in the fiducial profile taken from \citetalias{Wise2019}, and
that the limiting factor is the self-gravity of the gas in the core,
rather than the radiative feedback.

\subsubsection{Variation of SMS models}

We constructed a rapidly accreting super-giant protostellar model 
(\tabref{tab:fitting SE2}) based on the stellar evolution calculations in \citet{Hosokawa2013}.
We here consider other SMS models, discuss their differences, and
justify our adoption of the stellar model based on \citet{Hosokawa2013}.

A recent study by \citet{Bear2020} showed that using the code MESA (Modules for Experiments in Stellar Astrophysics; \citealt{Paxton2011}) non-accreting SMS radii can be an order of magnitude smaller than in \citet{Hosokawa2013} and \citet{Haemmerle2018}. 
The smaller radii are due to lack of entropy injection to stellar surface by rapid gas accretion, which is not included in the calculations of \citet{Bear2020}.
They also point out that if the central object is a fully convective object and represented by a polytrope of $n=3$, the stellar radii would be
also an order of magnitude smaller than in our case 
\citep[equation 2 in][]{Begelman2010}.
However, a rapidly accreting SMS, as considered in
this paper, is not well represented by a polytrope of $n=3$. 
Instead it has a compact massive core and an extended dilute
envelope \citep[e.g. fig. 2 of][]{Hosokawa2013}.
Thus, a stellar model which allows growth by accretion and includes entropy injection at the surface is more appropriate to adopt for the present study.

\subsection{Comparison to other works}
\subsubsection{Comparison to a PopIII formation case}
\label{subsec:1D-vs-2D}

The study presented in this paper is analogous to previous work, which
assessed, via radiation-hydrodynamical simulations, the final masses
of PopIII stars forming in primoridal gas that has cooled and contracted
inside lower-mass minihaloes~\citep{Hirano2015,Hirano2017}.  The two
major differences are that (i) here we consider proto-stellar cores and
their surrounding initial density profiles extracted from simulations
of more massive atomic-cooling haloes, and (ii) we assume spherical
symmetry, and perform 1D simulations, rather than the 2D treatment in
\citet{Hirano2015,Hirano2017}.

To check how our 1D simulation might give a different result compared
to a multidimensional simulation, we perform a 1D simulation as above,
but with the initial conditions adopted for a PopIII star forming
cloud from \citet{Hirano2015} - specifically their cloud ID=4 and with
$J_{21}=0$ (see their Table 1).  This 1D simulation can be directly
compared to the 2D simulation performed in their paper.  We find that
in our case, the cloud forms a star with a final mass of $\sim
2000\,\msun$, whereas in the 2D RHD simulation by \citet{Hirano2015}
the final mass is $\sim 50\,\msun$.  The difference can be attritubed
to the fact that in the 1D case, the cloud is spherical, and radiation
feedback is more easily suppressed. In the 2D simulations, radiation
can escape along the lower-density polar regions, and subsequently
ionise and heat the gas farther away, and more easily suppress the
accretion rates at these larger distances \citep{Tan2004,McKee2008}.  
This suggests that our 1D treatment may overestimate the final stellar mass.

\subsubsection{Comparison to 3-D simulations}

\citet{Luo2018} and \citet{Ardaneh2018} explored the early evolution of the direct collapse of protogalactic clouds with 3-D radiation hydrodynamical simulations. They 
used the flux-limited diffusion (FLD) approximation for computing the radiation flux assuming a grey opacity. They showed that the radiation luminosity emerging from the photosphere of the central core approaches the Eddington luminosity. 
The large luminosity affects the evolution of the cloud:  anisotropic recurrent outflows are driven by this strong radiation, as well as by thermal pressure, and disrupt
the central object.
The outflows collide with inflowing gas from larger scales, and are trapped.  However, they facilitate the outward transfer of angular momentum. As a result, $\sim 100\,\yr$ after formation of the photosphere, a rapidly accreting, quasi-spherical central object emerges, with no significant rotation.

Our work here is complementary, in that we follow the subsequent growth of the emergent protostar, and find that it can continue accreting without significant radiation feedback. 
The difference of the results may be attributed to the different
simulation times, scales, dimensions and treatment of a radiation field:
in our simulations, we consider a 1-D geometry, a simulation time up to
$3\,\Myr$, scales down to $0.003 \,\pc$ and multi-frequency radiation transfer 
while in \citet{Luo2018} and \citet{Ardaneh2018} they consider 3-D
geometry, a simulation time up to $\sim100\,\yr$ after formation of the photosphere, scales down to $10^{-7}\,\pc$, and a grey approximation for radiation transfer.

Specifically, if we resolved smaller scales in our simulations, an
\hii~region could expand in the early phases and cause feedback
(\secref{sec:resolution}) as seen in the 3-D simulations. 
Even in this case, the \hii~region would 
shrink in the later phase because the expected size of the \hii~region 
is much smaller than a protostar's Bondi radius. 
On the other hand, if we relax the spherical assumption, the feedback may also affect the later evolution of the flow. 
The difference of the grey and multi-frequency radiation transfer might also affects the result.

It is still unclear whether the early feedback as shown in 
\citet{Luo2018} and \citet{Ardaneh2018} continues to be important during
the later stages of evolution and on larger scales in the direct collapse
clouds.
To self-consistently explore long-term evolution of the clouds in the
small and large scales, multidimensional multi-frequency radiation
hydrodynamical simulations are awaited.

\subsection{Caveats}
\subsubsection{Spherical assumption}
\label{sec:spherical assumption}

%expected results if the spherical assumption is removed
Although we assume a spherically symmetric gas distribution in our
simulations, the gas distribution in the halo LWH of \citetalias{Wise2019} has
non-negligible angular momentum and an asymmetric morphology.  From
the Extended Data Figure~4 in \citetalias{Wise2019}, the circular velocity of the cloud
is comparable to the Keplerian velocity at radii where the enclosed
mass is $M_{\rm enc}\gtrsim2000\,\msun$, following a self-similar
solution of a gravitationally collapsing cloud as seen in the normal
Pop III star formation \citep{Yoshida2006} and direct-collapse of a
massive atomically-cooling gas \citep{Inayoshi2014b}.  If the
subsequent cloud evolution were followed by a multi-dimensional
simulation, the protostellar growth and the radiation feedback could
be changed for $M_*\gtrsim10^3\,\msun$.  For example, if an accretion
disc forms, and the density in the bipolar regions becomes low, the
stellar radiation can more easily break out of the inner regions.
While on small scales, this may help accretion in the shielded
equatorial plane, the radiation would become isotropic further out,
where the densities are lower, and could suppress the accretion at the
larger radii - as suggested by the direct comparison presented for one
case in \S~\ref{subsec:1D-vs-2D} above.

%effect of an accretion disc
More generally, protostellar evolution during an accretion phase is
more complex when an accretion disc forms.  A self-gravitating disc
can fragment via gravitational instability.  The fragments then fall
on to the central protostar and raise the accretion rate, making the
star bloat up and suppressing the ionising photon emissivity
\citep{Inayoshi2014, Sakurai2016b, Hosokawa2016, Tagawa2020,
  Chon2020}.

%shock formation
Furthermore, in a non-spherically symmetric morphology, supersonic
flows can form shocks (e.g., accretion shocks at the outer edge of a
rotationally supporter disc).  Shocks can then heat up the gas,
increasing the pressure, and possibly slowing down the infall.

%turbulence effect, effect of fixing the central star
\citet{Regan2020} recently performed high-resolution 3-D hydrodynamical
simulations of pristine atomic-cooling haloes to study the formation and
evolution of very massive stars. They found that the gas cloud in the core
of the halo is highly turbulent, and that the protostars are often in
low-density regions, accreting inefficiently. In our simulations we assume
that the central protostar is never displaced from the high-density region.
In this sense, we may overestimate the growth of the protostar.

%outflows and magnetic fields
In the context of present-day massive star formation, in addition to stellar radiation, collimated outflows and magnetic fields are also known to suppress stellar growth \citep{Cunningham2011,Kuiper2015,Kuiper2016,Rosen2020}.
Outflows suppress the stellar growth rate by sweeping up interstellar
material in polar directions of the star and ejecting the material 
from the star-forming system, as well as by decreasing the density in the 
polar directions and making stellar radiation feedback more effective. 
Strong magnetic fields also decelerate the growth rate since magnetic pressure slows down gravitational collapse of the cloud.
Magnetic fields also enhance angular momentum transport by magnetic
braking and inhibit the formation of a gravitationally unstable accretion 
disc which can cause fragmentation. These effects, however, may also 
suppress the accretion rate, rather than help the stellar growth \citep[Section 3.2 of][]{Rosen2020}.
The effects of outflows and magnetic fields in the SMS
formation case need to be investigated to clarify if they could be 
obstacles for the SMS formation.

%prospects
In order to more robustly judge whether the protostar emerging in the
core of a dynamically heated, atomic-cooling halo, can grow to a SMS,
multi-dimensional simulations will need to be performed, incorporating
the asymmetric distribution and nonzero angular momentum of the nearby
gas.

\subsubsection{Resolution of the simulations}
\label{sec:resolution}

We set the innermost cell radius to $r_{\rm min}=10^{16}\,\cm$, which is comparable to 
the initial protostar's gravitational radius $R_{\rm B}\sim 8.2\times 10^{15}\,\cm$, assuming
$T_\infty=300\,\K$ and $\mstar=2\,\msun$ (\secref{sec:Hydrodynamical simulations}). 
We compare this radius to the size of the \hii~region estimated from an
equilibrium St\"{o}mgren sphere, 
\begin{align}
\label{eq:Rstrom}
R_{\rm S}=\left(\frac{3Q_{\rm ion}}{4\pi n_{\rm e}^2\alpha}\right)^{1/3}, 
\end{align}
where $Q_{\rm ion}$ is the ionising photon emissivity and
$\alpha=2.6\times 10^{-13}(T_{\rm ion}/10^4\,\K)^{-0.85}\,{\rm cm^3\,s^{-1}}$ 
is the case-B recombination rate of hydrogen. 
We find $R_{\rm S}\sim 1.4\times 10^{13}\,\cm$ for $T_{\rm ion}=10^4\,\K$, $\mstar=5\,\msun$ and $Q_{\rm ion}\sim 10^{45}\,\s^{-1}$ \citep{Schaerer2002}, which is much smaller than either the gravitational influence radius or the resolution of our simulation.
If we estimate $R_{\rm S}$ assuming a density profile $\rho\propto r^{-1.5}$ instead of a constant density, $R_{\rm S}$ is even smaller, and becomes comparable to the initial stellar radius.
If we resolved a region as small as $r<R_{\rm S}$ in the simulations, 
an \hii~region may begin to drive an outflow and expand in the early phase. 
However, the \hii~region size (computed assuming the $\rho\propto r^{-1.5}$ profile)  is about four orders of magnitude smaller than
the growing protostar's Bondi radius.  
We conclude that even if the \hii-region is less compact due to an early outflow, it is unlikely to 
decelerate the inflow of neutral gas in the region
$R_{\rm S}\lesssim r \lesssim R_{\rm B}$ and radiation feedback would not be effective to hinder 
gas accretion \citep{Inayoshi2016, Sakurai2016}.
Even as the stellar masses grow during the evolution in our simulation, 
$R_{\rm S}$ remains below $R_{\rm B}$ by at least two orders of magnitude. 
Although numerical limitations preclude us from using a smaller minimum cell radius $r_{\rm min}$ and
resolving the initial ultra-compact \hii-region, we expect that our main result, i.e. that the 
radiation feedback is ineffective, is not compromised by this limitation.

%===============%
%==  Summary  ==%
%===============%
\section{Summary}
\label{sec:summary}

%summary of the results
\citetalias{Wise2019} argued, based on three-dimensional cosmological
simulations, that SMSs may form in large atomic-cooling haloes in
which H$_2$ molecules are not fully dissociated by external FUV
radiation.  In this work, we followed the evolution of a protostar
identified in one of the haloes (specifically, the halo ``LWH'' in
\citetalias{Wise2019}), beyond the point where their simulation stopped.
We performed 1-D radiation hydrodynamical simulations to explore if
radiation feedback suppresses the growth of this protostar.  We solved
the non-equilibrium chemical reactions of nine primordial species, and
included the radiation of the central source derived from stellar
evolution models, as well as radiation from a circumstellar disc.

We found that a SMS with a mass of $\gtrsim10^5\,\msun$ forms, even
though stellar radiation feedback temporarily halts the accretion for
$\sim10^4\,\yr$.  This feedback is caused by LW radiation from the
protostar.  The LW radiation dissociates H$_2$ in the inner region,
increasing the gas temperature and the gas pressure gradient force
which opposes gravity.  The feedback stops after $\sim10^4\,\yr$,
because the gas self-gravity and inward ram pressure force of the gas
building up on larger scales overcome the outward pressure gradient
force.  Although the stellar UV radiation is strong, no \hii~region
forms during the evolution because of the high density and efficient
hydrogen recombination.  We conclude that the protostar can grow to
$M_*\gtrsim10^5\,\msun$, as long as the central density is at least
$\sim$10\% of the value found in \citetalias{Wise2019}.   The main caveat
to this conclusion is our assumption of spherical symmetry; radiation
may have a stronger effect on an asymmetric collapse. Multi-dimensional
simulations will be required to include these effects and to assess
the robustness of our results.

%====================================================%
%==  Acknowledgements  ==================================%
%====================================================%
\section*{Acknowledgements}
The authors thank John Wise for fruitful discussions and comments and for 
providing data of the cloud profiles in \citetalias{Wise2019}.
The authors also thank Shingo Hirano for providing cloud profile data in
\citet{Hirano2015}.
This work is partially supported by Grant-in-Aid for JSPS Overseas
Research Fellowships (YS), NASA grant NNX17AL82G and NSF grant 1715661
(ZH), the National Science Foundation of China (11721303, 11991052,
11950410493; KI), and the National Key R\&D Program of China
(2016YFA0400702; KI).  The numerical simulations were partly performed
using services and resources provided by the Partnership for an
Advanced Computing Environment (PACE) at the Georgia Institute of
Technology, Atlanta, Georgia, USA.  The calculations were also carried
out in part on Cray XC50 at Center for Computational Astrophysics,
National Astronomical Observatory of Japan.
The data underlying this article will be shared on reasonable request to the corresponding author.

\small{\bibliography{ms}}

\end{document}